# Scalable and Stable Ferroelectric Non-Volatile Memory at > 500 °C

Dhiren K. Pradhan,[1,a] David C. Moore,[2,a] Gwangwoo Kim,[1] Yunfei He,[1] Pariasadat Musavigharavi,[1,3] Kwan-Ho Kim,[1] Nishant Sharma,[1] Zirun Han,[1,4] Xingyu Du,[1] Venkata S. Puli,[2] Eric A. Stach,[3] W. Joshua Kennedy,[2,*] Nicholas R. Glavin,[2,*] Roy H. Olsson III,[1,*] Deep Jariwala[1,*]

[1]Department of Electrical and Systems Engineering, University of Pennsylvania, Philadelphia, Pennsylvania, 19104, USA

[2]Materials and Manufacturing Directorate, Air Force Research Laboratory, Wright-Patterson AFB, OH, 45433, USA

[3]Department of Materials Science and Engineering, University of Pennsylvania, Philadelphia, Pennsylvania 19104, USA

[4]Department of Physics and Astronomy, University of Pennsylvania, Philadelphia, Pennsylvania 19104, USA

Non-volatile memory (NVM) devices that reliably operate at temperatures above 300 °C are currently non-existent and remains a critically unmet challenge in the development of high-temperature (T) resilient electronics, necessary for many emerging, complex computing and sensing in harsh environments. Ferroelectric $Al_xSc_{1-x}N$ exhibits strong potential for utilization in NVM devices operating at very high temperatures (> 500 °C) given its stable and high remnant polarization ($P_R$) above 100 μC/cm$^2$ with demonstrated ferroelectric transition temperature ($T_C$) > 1000 °C. Here, we demonstrate an $Al_{0.68}Sc_{0.32}N$ ferroelectric diode based NVM device that can reliably operate with clear ferroelectric switching up to 600 °C with distinguishable On and Off states. The coercive field ($E_C$) from the Pulsed I-V measurements is found to be -5.84 ($E_{C-}$) and +5.98 ($E_{C+}$) (+/- 0.1) MV/cm at room temperature (RT) and found to decrease with increasing temperature up to 600 °C. The devices exhibit high remnant polarizations (> 100 μC/cm$^2$) which are stable at high temperatures. At 500 °C, our devices show 1 million read cycles and stable On-Off ratio above 1 for > 6 hours. Finally, the operating voltages of our AlScN ferrodiodes are < 15 V at 600 °C which is well matched and compatible with Silicon Carbide (SiC) based high temperature logic technology, thereby making our demonstration a major step towards commercialization of NVM integrated high-T computers.

[a] Dhiren K. Pradhan and David C. Moore contributed equally to this work.

[*]Authors to whom correspondence should be addressed: william.kennedy.21@afrl.af.mil, nicholas.glavin.1@afrl.af.mil, rolsson@seas.upenn.edu, dmj@seas.upenn.edu.

**Introduction:**

Non-volatile memory (NVM) technologies that reliably operate at temperatures above 300 ºC are not available commercially and remain a critical gap in the design of extreme environment electronics.[1,2] There are numerous emerging harsh environment applications including aerospace, space exploration, oil and gas exploration, nuclear plants, mining and other applications that require complex computing and sensing capabilities in-situ.[3-7] Current Silicon (Si)-based micro(nano)electronics, utilizing complementary metal oxide semiconductor (CMOS) technology, encounter reliability challenges above 200 °C and cannot retain their functional properties at high temperatures as the number of thermally induced carriers exceeds the doping concentration level.[6,8] Extensive research has led to well-established wide-band gap semiconductor material solutions e.g. Silicon Carbide (SiC) for logic transistors, which can effectively operate at temperatures as high as 800 °C,[3,9-11] but the absence of NVM devices suitable for information storage hinders the execution of intricate computing operations at these elevated temperatures.

Incumbent nonvolatile random-access memory (NVRAM) technologies relying on magnetic, FLASH, phase-change, and resistive switching mechanisms rapidly degrade at < 300 °C.[12-15] SiC electronics, being the most mature and preferred high-temperature logic platform, further emphasizes the importance of this pursuit.[11] Microelectromechanical system-based flash memory (MEM-FLASH) and nanogap resistance switching (NGS) technologies show promise for NVM applications at high temperatures.[8,16] However, these approaches have drawbacks due to the presence of moving parts and the challenges associated with operating in diverse atmospheres. Among the NVM technologies explored at temperatures exceeding 300 °C, three have received significant attention: flash memory, resistive memory based on oxides, and ferroelectric memory.[17-20] Among these, only flash memory, which operates at approximately 200 °C, is a commercially available product.[1] Therefore, active research and development is necessary to produce stable and scalable NVM devices at temperatures > 200 °C while ensuring voltage and process compatibility with SiC logic, which is the most mature transistor technology capable of high-temperature operation.

Ferroelectric materials exhibit remanent polarization, field-driven switching, and the ability to retain the switched state for an extended time. This makes them well-suited for low-power nonvolatile memory applications as they also exhibit fast switching speed, low switching energy, and have potential for multibit operation.[21-23] To meet the temperature demands of harsh environment applications, ferroelectric materials with high Curie temperatures ($T_C$) are required as ferroelectric polarization and reliable switching cannot be observed if the device temperature exceeds the $T_C$.[24] While numerous works on perovskite oxide and hafnia/zirconia based fluorite oxide ferroelectric materials have shown ferroelectricity at high temperatures none have been able to demonstrate stable and scalable NVM device operation at 500 °C. This is mainly due to material drawbacks such as high atomic diffusivity, low-coercive fields, low-remnant polarizations and lower ferroelectric $T_C$ of the perovskite and fluorite oxide ferroelectric materials.

In contrast wurtzite structured III-Nitride based ferroelectric $Al_xSc_{1-x}N$ materials have strong potential to be utilized in NVM devices, which can operate at very high temperatures as they concurrently exhibits high $P_R > 100$ μC/cm$^2$, large coercive field ($E_C$) >2 MV/cm, and a very high ferroelectric transition temperature ($T_C$) > 1000 °C.[23] $Al_{0.68}Sc_{0.32}N$-based scaled ferroelectric memory devices have already been demonstrated in multiple reports with robust ferroelectricity even demonstrated in 5 nm thick films.[21-23,25,26] Although in-situ measurements were not performed, it was observed that $Al_xSc_{1-x}N$ thin films with thicknesses exceeding 400 nm, grown on molybdenum bottom electrodes, exhibited structural and polarization stability at temperatures as high as 1025 °C.[27,28] Minimal changes in the remanent polarization values has been observed up to 400 °C for 225 nm thick ferroelectric $Al_{0.70}Sc_{0.30}N$ film on a Pt (111) bottom electrode.[29] The existence of stable and large $P_R$ at elevated temperature (> 400 °C) and thermal stability of $Al_xSc_{1-x}N$ up to 1100 °C make this a promising avenue for storing information in computing applications in harsh environments (> 500 °C). [28-30]

Here, we demonstrate a 45 nm thick $Al_{0.68}Sc_{0.32}N$ ferroelectric diode based NVM device that can operate up to 600 °C at ≤ 20 V operation, ensuring compatibility with high T SiC logic devices.[9-11] The devices are composed of metal-insulator-metal (MIM) structures of Ni/$Al_{0.68}Sc_{0.32}N$ (45 nm)/Pt (111) grown on 4" Silicon wafers. We report detailed temperature dependent ferroelectric and NVM characteristics up to 600 °C in-situ. The temperature dependent DC I-V curves exhibit ferroelectric diode behavior with clear ferroelectric switching up to 600 °C with distinguishable ON and OFF states. At 500 °C, these devices show an unprecedented 1 million read cycles and stable On-Off ratio for > 6 hours.

**Results and Discussion:**

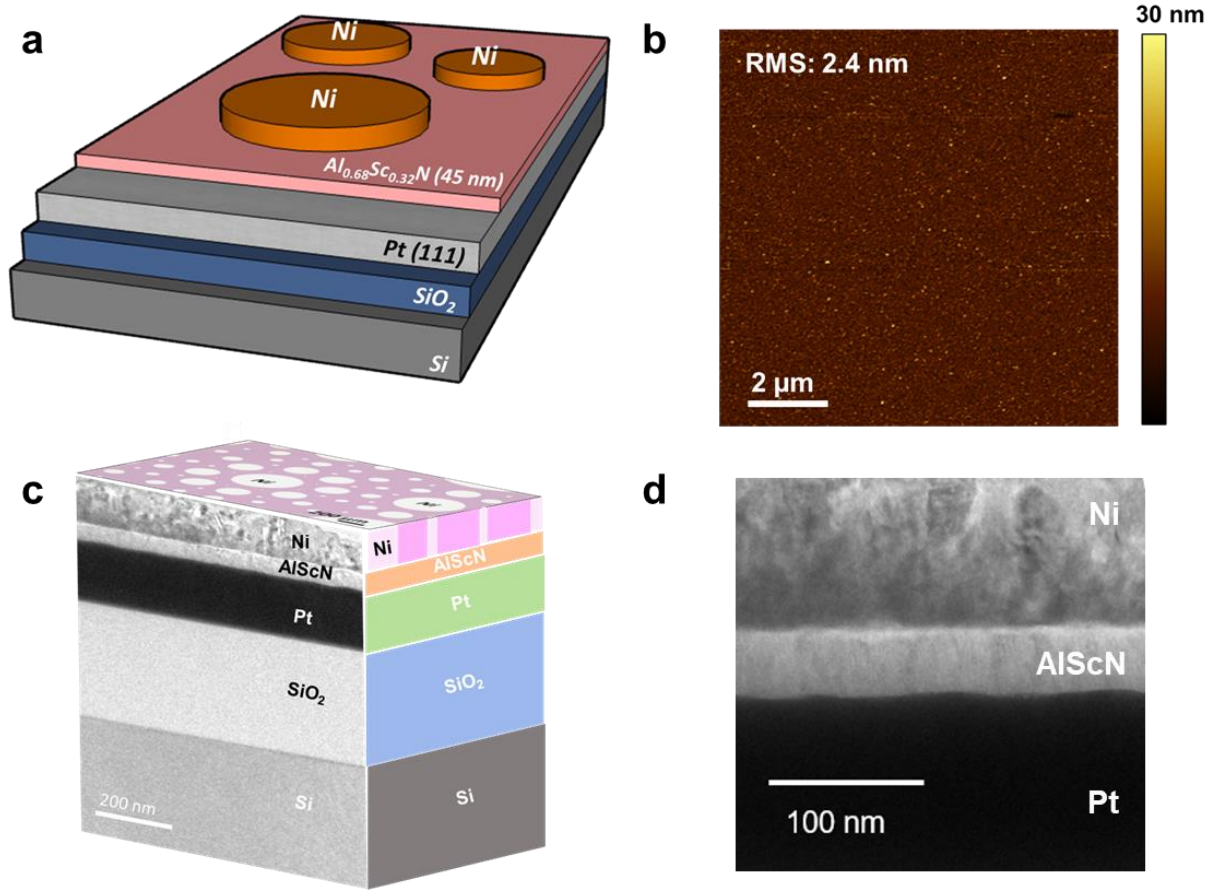

Figure 1. (a). Schematic (b) AFM height image (c) a composite image of the AlScN device showing each layer (on the left is cross sectional TEM and on the right is a labeled schematic and on top is an optical microscopic image) (d) cross sectional TEM image of the Ni/Al$_{0.68}$Sc$_{0.32}$N/Pt (111) MIM device.

A schematic depiction of the Ni/Al$_{0.68}$Sc$_{0.32}$N/Pt NVM devices investigated in this study is shown in Fig. 1(a). First, a 45 nm thick Al$_{0.68}$Sc$_{0.32}$N film was deposited by co-sputtering from separate Al and Sc targets onto a commercially available 150 nm Pt(111)/Ti/SiO$_2$ on Si(100) substrate.[21,22] Top electrodes (TE) were then patterned using standard photolithography, followed by sputtering of Ni (100 nm) metal top electrodes and a lift-off process. Circular top Ni electrodes of various sizes were patterned on the AlScN surface (further details in supplementary Fig. S1). Fig. 1(b) presents the surface morphology of the top AlScN layer in the Ni/AlScN/Pt device. Surface topography reveals a smooth and uniform surface with a root mean squared roughness ($R_{rms}$) of ~ 2.4 nm. The heterostructure's upper surface exhibits nano-scale smoothness and homogeneity, devoid of any microcracks, pores, or holes. A composite image of the AlScN MIM device displaying each layer with a cross sectional TEM image on the left and a labeled schematic on the right with an overlaying optical microscopic image of the top surface of AlScN. A cross sectional TEM image of the MIM device is shown in Fig. 1(d). The TEM image confirms the existence of all the layers and their respective thicknesses. The thickness of the $Al_{0.68}Sc_{0.32}N$ layer was found to be ~ 45 (±2) nm. A native oxide layer of thickness ~ 4 nm was observed on the surface of AlScN film as the films were exposed to air prior to top Ni electrode deposition. The cross-sectional TEM image recorded at high resolution is also shown in Fig. S2 (further details in supplementary Fig. S2).

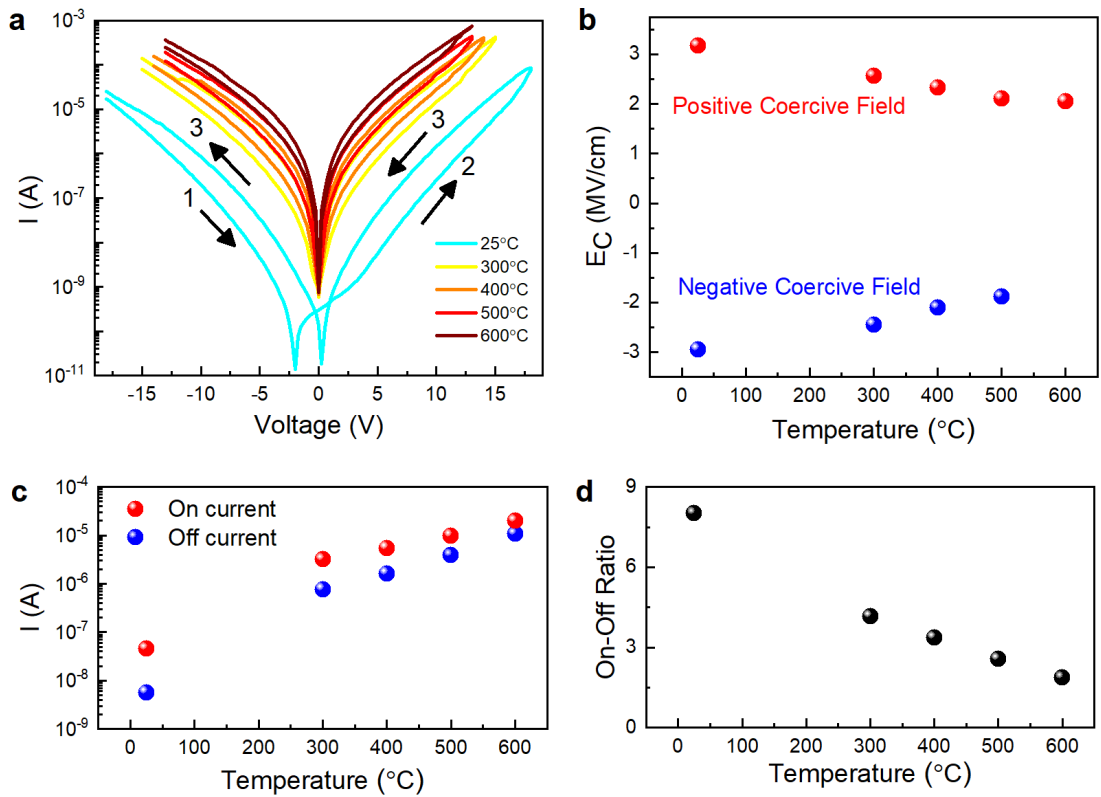

Figure 2. (a) Quasi-DC current (I) versus voltage (V) at selected temperatures (b) Coercive Field versus temperature (c) On and Off current (d) On-Off ratio at 5 V read voltage as a function of temperature for the Ni/Al$_{0.68}$Sc$_{0.32}$N/Pt (111) MIM device.

To investigate the impact of temperature on ferroelectric diode characteristics, ferroelectric switching, and leakage current behavior, quasi-DC (0.01 Hz) I–V hysteresis measurements were performed at different temperatures (25-600 °C) on the MIM devices (Fig. 2(a)). Upon a positive voltage sweep (1 to 2: black arrows in the figure), the device undergoes a transition from high-to-low resistance state, accompanied by a polarity shift from a negative-forward diode to a positive-forward diode. Likewise, during a negative voltage sweep, the device exhibits a polarity change from a positive-forward diode to a negative-forward diode. At all temperatures, ferroelectric diode-like behavior can be clearly observed in the Al$_{0.68}$Sc$_{0.32}$N devices from the change in the resistance upon ferroelectric switching. Positive coercive field ($E_{C+}$) and negative coercive field ($E_{C-}$) values were calculated using the first derivative of current with respect to voltage (further details in supplementary Figs. S3 and S4) and their temperature dependence of $E_{C+}$ and $E_{C-}$ is depicted in Fig. 2 (b). The $E_{C+}$ values at room temperature (RT, ~25 °C) and at 600 °C were found to be 3.16 and 2.05 (+/- 0.1) MV/cm, respectively. The $E_{C-}$ values at RT and at 500 °C were observed to be -2.94 and -1.88 (+/- 0.1) MV/cm, respectively. Note that the $E_{C-}$ value at 600 °C is not shown Fig. 2 (b) as it is inconclusive. Both $E_{C+}$ and $E_{C-}$ decrease linearly with increasing temperature, indicating the increased thermal energy overcomes the activation barrier and facilitates the crystal dipole transformation associated with domain switching, matching observations previously reported in similar devices.[29,31] The reduction of $E_C$ at high temperatures can also be correlated to the increased thermally activated domain wall motion.[32] This appears to correlate with the Curie-Weiss law far away from critical behavior as has been reported in other ferroelectric materials,[33,34] suggesting that the devices would function well above 600 °C, which is the maximum temperature of the probe station used in this work. To further corroborate the ferroelectric behavior of the MIM devices the voltage dependence of capacitance (C–V) measurement was carried out and it exhibits a butterfly shaped loop indicating a non-linear capacitor. The decrease of capacitance with increasing of applied voltage, demonstrating ferroelectric polarization switching (see further details in supplementary Fig. S5) is observed.

Fig. 2 (c) shows the temperature dependence of On and Off currents at 5V read voltage, with both the On and Off currents increasing with rising temperature. The On-Off ratio of the AlScN MIM devices at 5V read voltage was calculated, and its temperature dependence is presented in Fig. 2 (d). The device exhibited an On-Off ratio of 8.06 at RT and 1.87 at 600 °C, and it decreases with increasing temperature. This reduction in On-Off ratio primarily results from an increase in overall leakage current arising from increasing thermal energy which reduces activation barrier for hopping or defect assisted tunneling transport in the AlScN.[29] De-trapping of charges owing to domain wall motion and relaxation of the AlScN crystal lattice around defect sites may also contribute to an overall increase in current at high temperatures. A theoretical analysis based on the Poole-Frenkel emission model regarding the decrease of the On-Off ratio with increasing temperature is included in the supplementary material in Fig. S6. Regardless, while the engineering of devices with enhanced On-Off ratios is the subject of an ongoing investigation and will be the subject of a follow up work, the absolute current difference does not diminish with increase of temperature (further details in supplementary Fig. S7). We suggest that a practical MIM device integrated with logic transistors could utilize a reference capacitor in the Off state to subtract a baseline and read successfully at all temperatures between RT and 600 °C.

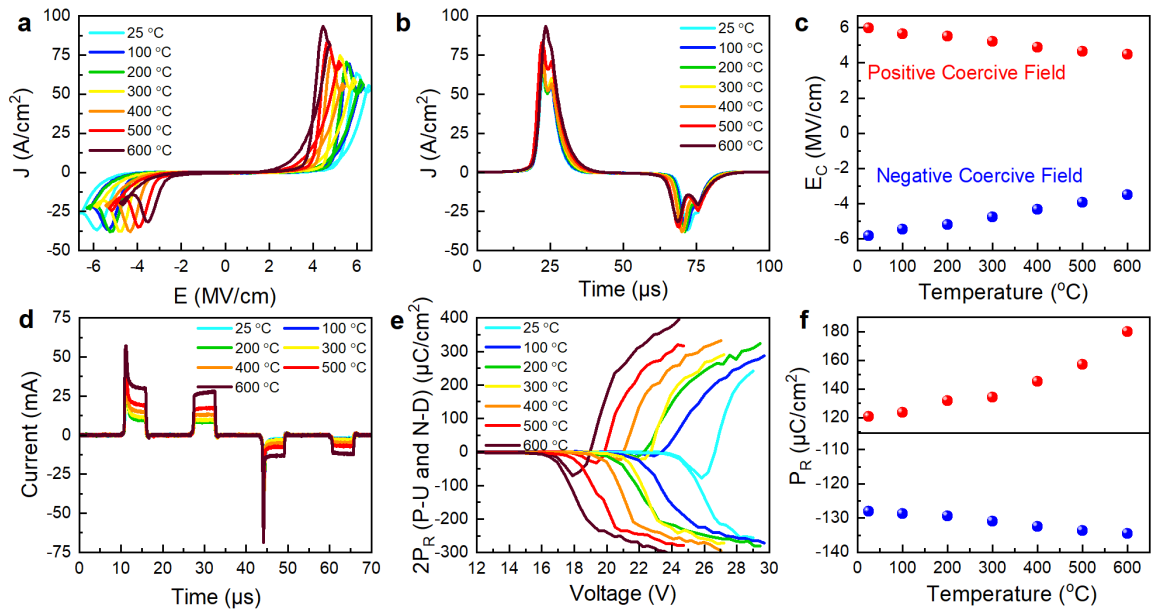

Figure 3. (a) Pulsed IV current density (J) versus electric field (E) at different temperatures (b) Pulsed IV current density (J) versus time (c) Coercive field ($E_C$) vs temperature (d) PUND traces at select temperatures (e) Remnant polarization vs PUND voltage for each temperature and (f) Remnant polarization vs. temperature measured at $E_C * 1.26$ for relative consistency. Note that positive $P_R$ values at 300 °C and beyond are inconclusive due to the large leakage currents observed at high temperatures.

Pulsed-IV, in contrast to quasi DC-IV, provides a dynamic characterization approach to investigate the electrical responses of ferroelectric devices under high voltage ramp rates, allowing the differentiation of ferroelectric switching dynamics from leakage and capacitive charging contributions. We performed pulsed-IV measurements across a temperature range from RT (~25 °C) to 600 °C in 100 °C increments. We used 10 kHz excitation signals on 100 µm diameter devices, incrementally increasing the excitation voltage until the device failed. At this ramp rate, leakage current was negligible compared to ferroelectric switching and capacitive charging currents until one reaches high voltages comparable to dielectric breakdown voltages, enabling direct observation of the switching voltage. Fig. 3(a and b) depicts selected traces before the onset of large leakage current, displaying the $E_C$ as the maximum and minimum values for on and off switching, respectively. The $E_{C+}$ value at RT and at 600 °C are found to be 5.98 and 4.47 (+/- 0.1) MV/cm, respectively whereas the $E_{C-}$ values at RT and at 600 °C are observed to be -5.83 and -3.50 (+/- 0.1) MV/cm, respectively. These values are well matched with the previous reports.[21,22,30] The peak current calculated from Fig. 3 (b) as a function of temperature is presented in Fig. S8. As has previously been observed with similar devices[35] apparent $E_C$ increased with ramp rate (further details with various ramp rates are available in supplementary Fig. S9). Fig. 3(c) presents the measured $E_{C+}$ and $E_{C-}$ plotted against temperature, values derived from the pulsed-IV as shown in in Fig. S10 with both exhibiting a linear reduction with respect to temperature as were observed with the DC-IV.

Positive-Up Negative-Down (PUND) measurements offer a time-resolved approach to probe the signature of ferroelectricity and to explore the polarization dynamics of AlScN ferroelectric devices, allowing observation of the ferroelectric response separated from leakage and capacitive charging.[36] We performed PUND measurements at each temperature, starting at 5V pulse amplitude, and increasing amplitude until device failure. The PUND measurement with voltage pulses and the current traces performed at RT are shown in Fig. S11. Fig. 3(d) illustrates select traces that clearly demonstrate ferroelectric switching for both positive and negative voltage pulses up to 600 °C. Given that the onset voltage for switching and leakage both vary with temperature, we compare the PUND results using pulse voltages normalized to the onset of diode behavior, defined as the point where the pulse current becomes three times larger than the capacitive charging current as shown in Fig. S12. PUND traces selected as "saturated" have maximum voltages of 126% of this onset value for each temperature. These onset voltages were found to be consistent with the pulsed-IV results, dropping approximately 1.3 V per 100 °C increase. We chose 126% to create the most robust and comprehensive dataset, as this voltage represents the maximum similar saturation available at all temperatures, as the devices do not fully saturate before device failure. Fig. 3(e) illustrates the switched polarization ($2P_R$) traces at all voltages and temperatures, with each trace ending at the point of device failure. Fig. 3(f) plots the remnant polarization versus temperature for both the P-U and N-D pulses. The positive and negative $P_R$ values were found to be 120 and -128 μC/cm² at room temperature that matches well with the previous reports.[21,23] As predicted, the magnitudes of both increased slightly with temperature due to the rise in overall leakage current, with the positive $P_R$ values being more affected by leakage at 300 °C and beyond making the positive $P_R$ values unreliable at elevated temperature.[31]

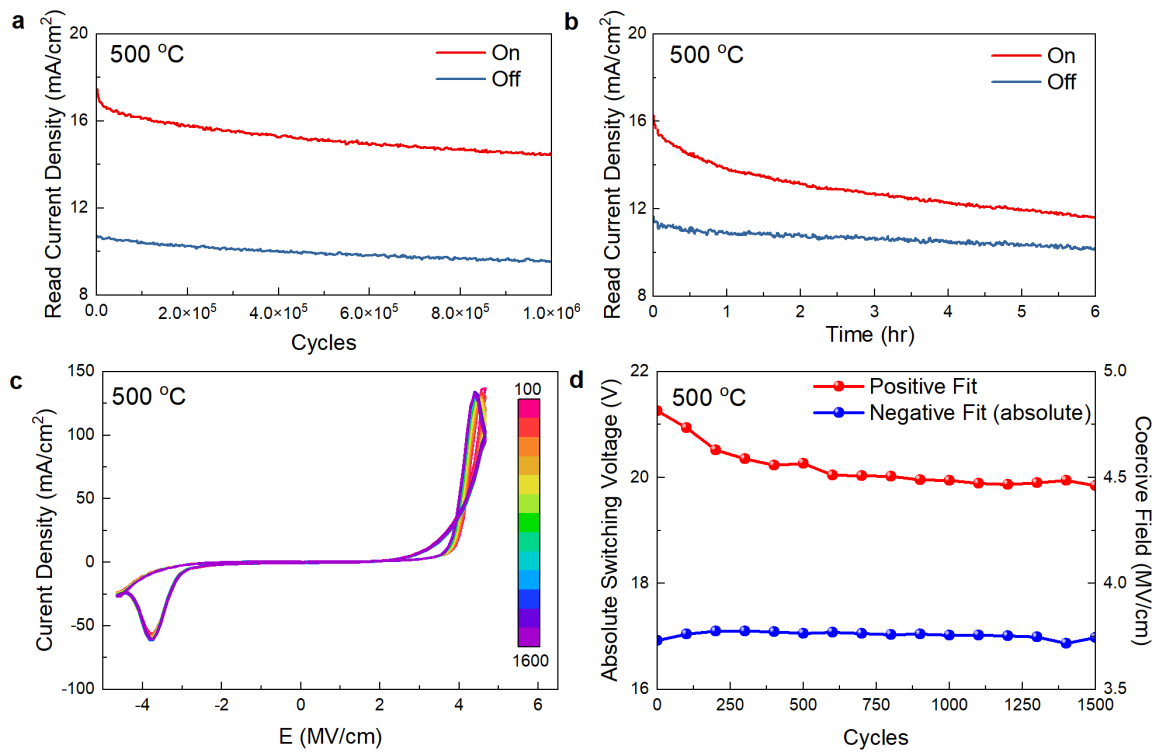

Figure 4. (a) >1M read endurance cycles at 5 V read voltage 500 °C (b) Retention behavior at 5 V read voltage at 500 °C (c) Pulse IV survival tests during 19 V write cycling at 500 °C. (d) Positive and negative switching voltages from test pulse IV vs. number of write cycles.

While the capacitors breakdown near the large fields required for writing, for a FE diode read operations can be achieved at low voltages that do not induce failure. Reads taken at 5V yield a current density of 10-17 mA/cm$^2$, providing a current of 3-6 µA for 100 µm diameter pads, depending on the device's polarization state. This was sufficient for measurement without stressing the device. The evolution of these currents over cycles and time is shown in Fig. 4(a) and (b), respectively, at 500 °C (representative current vs. time data available in Fig S13). The 1M read retention measurement was performed with separate runs for the On and Off devices due to timing restrictions on the DAQ. Greater than 1M cycles of 100 µs width were successfully read over the course of approximately two minutes, showing a modest decrease in the On-state current within the first 20k cycles, followed by a slight continuing decrease in both On and Off-state current density. We attribute this slope to AlScN self-annealing at high temperature, though the higher drop in the On-state device likely indicates heightened domain relaxation. The 6-hour retention test (Fig. 4(b)) was performed simultaneously with On and Off devices, but due to limitations in the hardware, a several-minute gap occurred between the write state and the read tests. A cluster of 25 consecutive 5 V pulses of 100 µs width were performed every 15 seconds and averaged to produce the current data. The On-state current continues to drop throughout the test, reducing the On-Off ratio to only 1.15 by the end, but a detectable gap is still evident. Improvements in the On-Off ratio and retention are subjects of ongoing research. However, even at this level and with the use of a reference pad, the state remains readable for logic device purposes. The On-Off ratio at RT was found to be ~ 8 and the retention behavior at RT is presented in Fig. S14.

Write longevity (Fig. 4(c) and (d)) was found to be sensitive to both write voltage and increasing temperature. Write endurance tests were performed in groups of 100 write pulses, followed by a test pulsed-IV to determine whether the device survived and observe any changes in device properties. At RT on 100 µm diameter pads, 25 V write pulses survived for 6033 ± 379 cycles (further details in supplementary Fig. S15). At 500 °C, devices survived 1433 ± 413 write cycles at 19V (Fig 4(c)) and 1233 ± 58 cycles at 20V write voltage. The $E_C$ values for the positive switching were observed to shift more than the negative $E_C$ during the write endurance testing, decreasing from 4.72 MV/cm (21.25 V) to 4.41 MV/cm (19.85 V) throughout the run, matching previous literature.[29] The wakeup effect is observed in the first few cycles(further details in supplementary Fig. S16), but the high-temperature long-term $E_{C+}$ drift is much higher than observed at RT (see supplementary Fig. S16) and has a reverse trend at the onset of $E_{C-}$, which may be due to repeated cycles inducing domain wall motion which de-pins domains.[37] However, no evident change occurred just before failure, and no obvious sign of impending failure is observed in the previous pulse. The precise mechanisms underlying the failure of these devices at high voltage and after numerous cycles, as well as strategies to improve their lifespan, remain the subject of ongoing investigation. The failure mode consistently involves a dielectric breakdown leading to an unrecoverable short circuit between the top and bottom pad and can be induced by overvoltage or excessive pressure from the probe tip on the top pad. Breakdown has been observed at both positive and negative voltages, typically above the switching voltage. Unlike current, the failure threshold varies significantly from one device to another, though the measurements depicted in Figs. 3 and 4 are representative. Overall, smaller devices appear to be more resilient, likely due to the decreased number of defects within their AlScN layer, reducing the opportunities for failure. This dielectric breakdown has been observed and reported previously by us[25] and literature elsewhere.[31]

**Conclusions:**

We demonstrate ferroelectric diode based NVM devices with an MIM structure of Ni/native oxide/Al$_{0.68}$Sc$_{0.32}$N/Pt (111) that can stably operate up to 600 °C. The temperature dependent ferroelectric properties and memory device characteristics were evaluated using quasi-DC, AC and pulsed IV measurements up to 600 °C. The temperature-dependent quasi-DC I-V curves exhibit clear ferroelectric switching up to 600 °C with distinct On and Off states. At 500 °C, these devices exhibit an unprecedented 1 million read cycles and maintain a stable On-Off ratio for over 6 hours. At elevated temperatures the write voltages of the devices remain below 15 V, demonstrating compatibility with SiC based logic devices. The devices demonstrated herein therefore represent a significant step towards realization of digital computing systems capable of operating at high temperatures and in a variety of harsh environment applications.

**Methods**

**Growth of Al$_{0.68}$Sc$_{0.32}$N**

The AlScN thin films containing 32% Sc concentration (Al$_{0.68}$Sc$_{0.32}$N) were grown on (111)-oriented Pt wafers using a physical vapor deposition (PVD) system. The growth conditions, substrates and some aspects of device fabrication are identical to our prior published reports.[21,22,25,38]

Following the deposition of the Al$_{0.68}$Sc$_{0.32}$N film, a 50 nm thick Al capping layer was deposited on the top of it. This deposition was performed at a temperature of 150 °C with an Ar gas flow of 20 sccm. It is worth noting that the deposition of the Al capping layer was done without breaking the vacuum, ensuring prevention of surface oxidation on the Al$_{0.68}$Sc$_{0.32}$N film.

**Device Fabrication**

To create the pattern for the top electrode, the Al capping layer underwent etching using a 1% HF solution. Subsequently, the $Al_{0.68}Sc_{0.32}N$/Pt (111) samples were coated with a negative photoresist (NR71-3000p), followed by photolithography to define the top electrode pattern. After development using a RD6 developer, a 100 nm thick layer of Ni was deposited as the top electrode metal using a sputtering system (Kurt J. Lesker PVD 75 PRO-Line Sputterer), The deposition rate was set at 2.5 Å/s and took place under low pressure condition of $5 \times 10^{-7}$ Torr. Then, the samples were immersed in a Remover PG (Photoresist Remover) for approximately 10 min, gently shaken to lift off any excess metal and then rinsed with deionized water. Finally, the samples were dried using $N_2$ blowing.

**AFM and TEM characterizations.**

To study the surface morphology of the top layer of $Al_{0.68}Sc_{0.32}N$ MIM device, AFM topography scans were performed on the AlScN sample using an OmegaScope Smart SPM (AIST-NT) setup. Additionally, scanning TEM characterization and image acquisition were conducted using a JEOL F200 instrument operated at 200 kV acceleration voltage. A cross-sectional TEM sample was prepared using a TESCAN S8000X system. Pt protecting layers were deposited using electron-beam and ion-beam techniques to ensure the preservation of the sample top surfaces and to prevent heating effects during focused-ion-beam milling. The focused-ion-beam milling of the lamella was carried out at 30 keV. Subsequently, an in-situ lift-out technique was used with the assistance of a Kleindiek probe manipulator. The final steps involved thinning and cleaning the lamella at energies of 10 keV and 5 keV, respectively.

**Device Characterizations**

In order to characterize the device performance, several electrical measurements including current-voltage (I-V), capacitance-voltage (C-V) and positive-up and negative-down (PUND) measurements were conducted at room temperature. The PUND test involved a voltage waveform consisting of four monopolar pulses. The pulse width, rise/fall times, and delay (interval between subsequent pulses) were specified as 5 μs, 800 ns, and 10 μs, respectively. The RT measurements were performed in air in a Cascade Microtech (MPS-150) probe station using a Keithley 4200A semiconductor characterization system. For temperature-dependent measurements such as DC I-V, pulsed I-V, PUND, the device was placed in a vacuum environment of approximately $10^{-4}$ Torr.

These measurements were conducted in a custom MicroXact high temperature probe station using tungsten probes. PUND and DC-IV characterization was performed using the same Keithley 4200A system. Pulsed I-V and write longevity measurements were performed with a Radiant Technologies Multiferroic II Ferroelectric Tester at 10 kHz unless otherwise specified. Read endurance measurements at RT were carried out by a Keithley 4200A semiconductor characterization system with a ±19 V DC pre-soak, followed by 5V, 100 μs square pulses for the read current. Similarly, the 500 °C read endurance measurements were performed with a Keysight B1500A Semiconductor Parameter Analyzer to accommodate limitations in the high temperature probe station consisted of simultaneously measuring On and Off electrodes with a ±13 V DC pre-soak, followed by 5V, 100 μs square pulses for the read current. Read retention measurements were performed with a Measurement Computing USB-1808X DAQ with the 5V, 10 kHz waveform provided by the Keysight B1500A on electrodes switched in the same manner as the read endurance measurements. All electrical measurements were carried out on top Ni circular electrodes with a 100 μm diameter unless otherwise specified, which were positioned on the $Al_{0.68}Sc_{0.32}N$/Pt samples. Maximum PUND, I-V, and write voltages were varied with changing temperature to account for temperature-variable switching voltage.

**Acknowledgements**

D.J., R.H.O., D.P. and G.K. acknowledge primary support for this work from AFRL via the FAST program. Z.H. acknowledges support from the VIPER program of the Vagelos Institute for Energy Science and Technology at Penn. V.S.P. acknowledges support for this work from AFRL via the National Research Council (NRC) senior research associate fellowship program. A portion of the sample fabrication, assembly, and characterization were carried out at the Singh Center for Nanotechnology at the University of Pennsylvania, which is supported by the National Science Foundation (NSF) National Nanotechnology Coordinated Infrastructure Program grant NNCI-1542153.

**Author contributions**

D.J., R.H.O., D.P. and G.K. conceived the devices, measurements and sample fabrication idea/concepts. D.P. and G.K. fabricated the samples with assistance from X.D. and measured them at room temperature with assistance from Y.H., N.S. and K.K. D.P. and D.M. performed all the high-temperature measurements with assistance from V.S.P. D.P. and D.M. analyzed all the electrical data. Z.H. performed the theoretical fits to the high-temperature I-V data. P.M. performed electron microscopy and data analysis under supervision of E.A.S. D.J., R.H.O, N.G and W.J.K. supervised and guided the project. D.P. and D.M. wrote the manuscript. All authors provided their inputs to the manuscript and supplementary information.

**Data availability**

All data are available in the paper and Supplementary Information are available from the corresponding authors upon reasonable request.

**Competing Interests Statement**

D.J., R.H.O., D.P., G.K. and Y.H. have a provisional patent filed on this work. The authors declare no other competing interests.

# Supporting Information

**Scalable and Stable Ferroelectric Non-Volatile Memory at > 500 °C**

Dhiren K. Pradhan,[1,a] David C. Moore,[2,a] Gwangwoo Kim,[1] Yunfei He,[1] , Pariasadat Musavigharavi,[1,3] Kwan-Ho Kim,[1] Nishant Sharma,[1] Zirun Han,[1,4] Xingyu Du,[1] Venkata S. Puli,[2] Eric A. Stach,[3] W. Joshua Kennedy,[2,*] Nicholas R. Glavin,[2,*] Roy H. Olsson III,[1,*] Deep Jariwala[1,*]

[1]Department of Electrical and Systems Engineering, University of Pennsylvania, Philadelphia, Pennsylvania, 19104, USA

[2]Materials and Manufacturing Directorate, Air Force Research Laboratory, Wright-Patterson AFB, OH, 45433, USA

[3]Department of Materials Science and Engineering, University of Pennsylvania, Philadelphia, Pennsylvania 19104, USA

[4]Department of Physics and Astronomy, University of Pennsylvania, Philadelphia, Pennsylvania 19104, USA

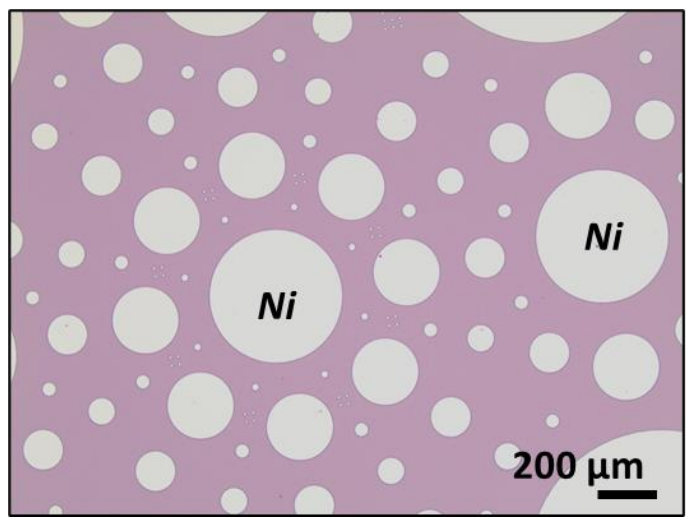

Fig S1. Optical microscopic image of the Ni/Al$_{0.68}$Sc$_{0.32}$N/Pt (111) MIM device. Circular top Ni electrodes (gray) of various sizes were patterned on the AlScN (pink) surface. The electrical measurements were carried out on 100 and 150 µm diameter top Ni electrodes.

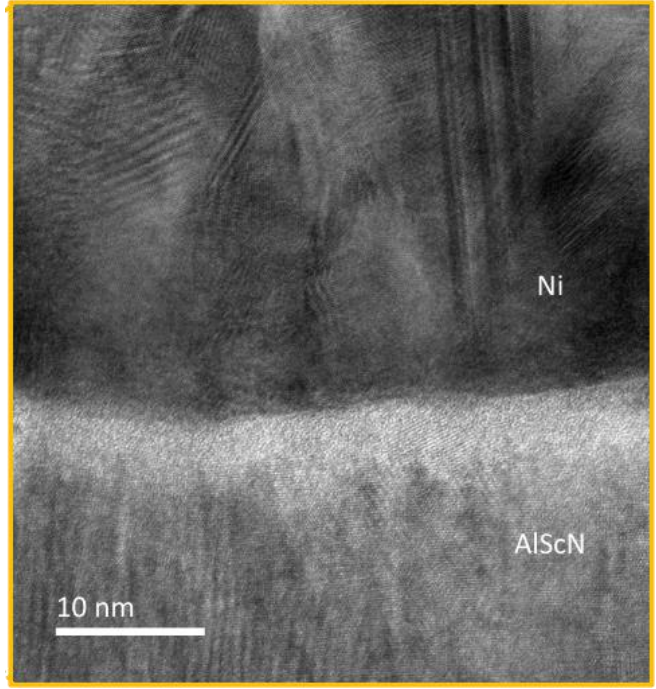

Fig S2. The cross-sectional TEM image of the Ni/AlScN interface of the Ni/Al$_{0.68}$Sc$_{0.32}$N/Pt (111) MIM device recorded at higher resolution.

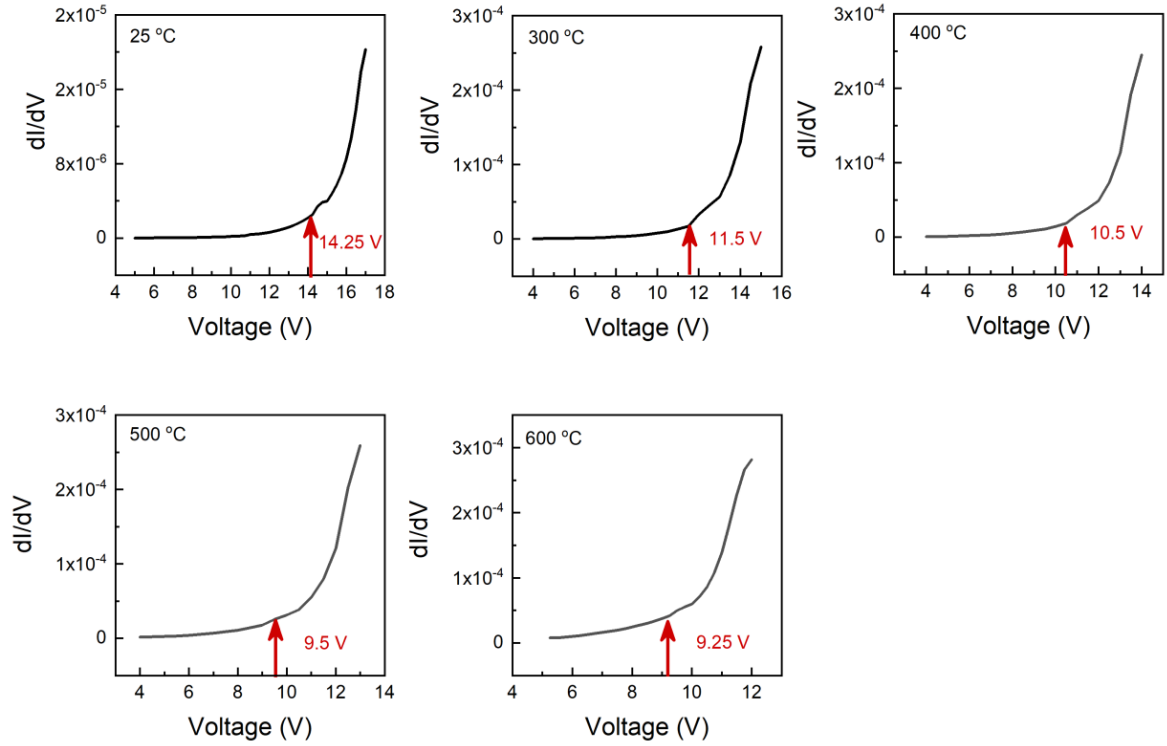

Fig. S3. Positive coercive voltage calculated from the first derivative of current with respect to voltage at different temperatures.
The positive coercive field decreases with increasing temperature.

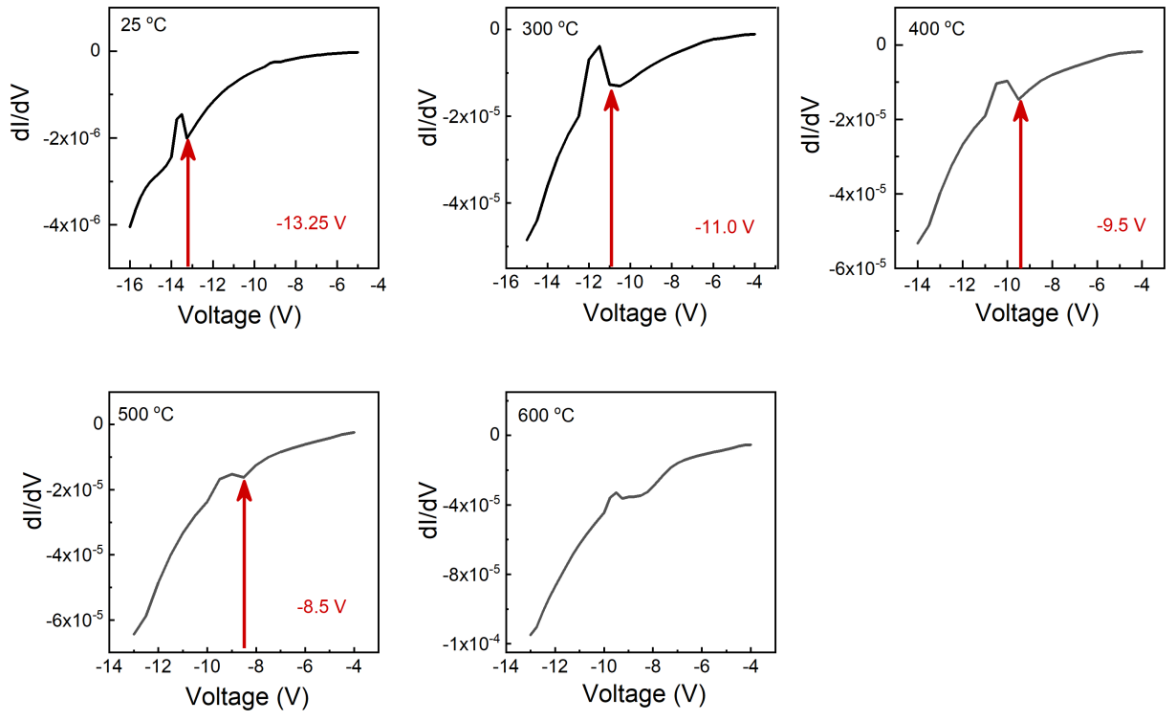

Fig. S4. Negative coercive voltage calculated from the first derivative of current with respect to voltage at different temperatures.

The negative coercive field decreases with increasing temperature. Note that the negative coercive field value at 600 °C is not shown Fig. 2 (b) as it is inconclusive.

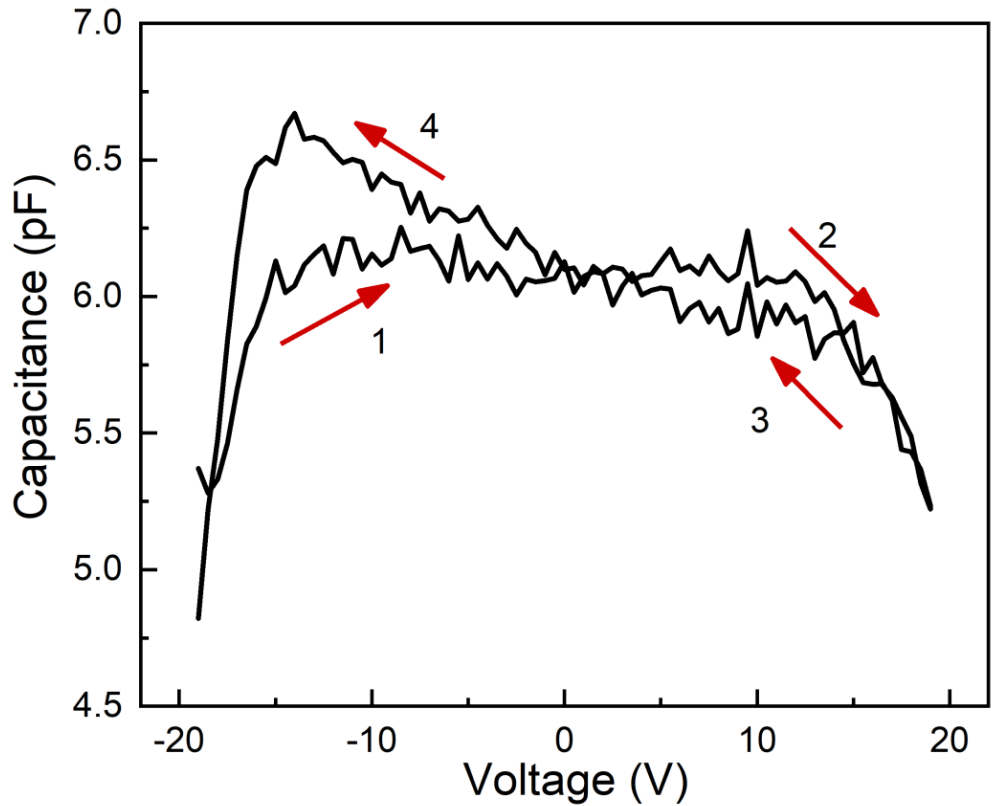

Fig. S5. C–V curve of the Ni/Al$_{0.68}$Sc$_{0.32}$N/Pt (111) MIM device measured room temperature illustrates a typical butterfly loop.

The butterfly-shaped loop observed in the C-V curve suggests the presence of a non-linear capacitor with decreasing capacitance as the applied voltage increases, which in turn indicates the occurrence of ferroelectric polarization switching.

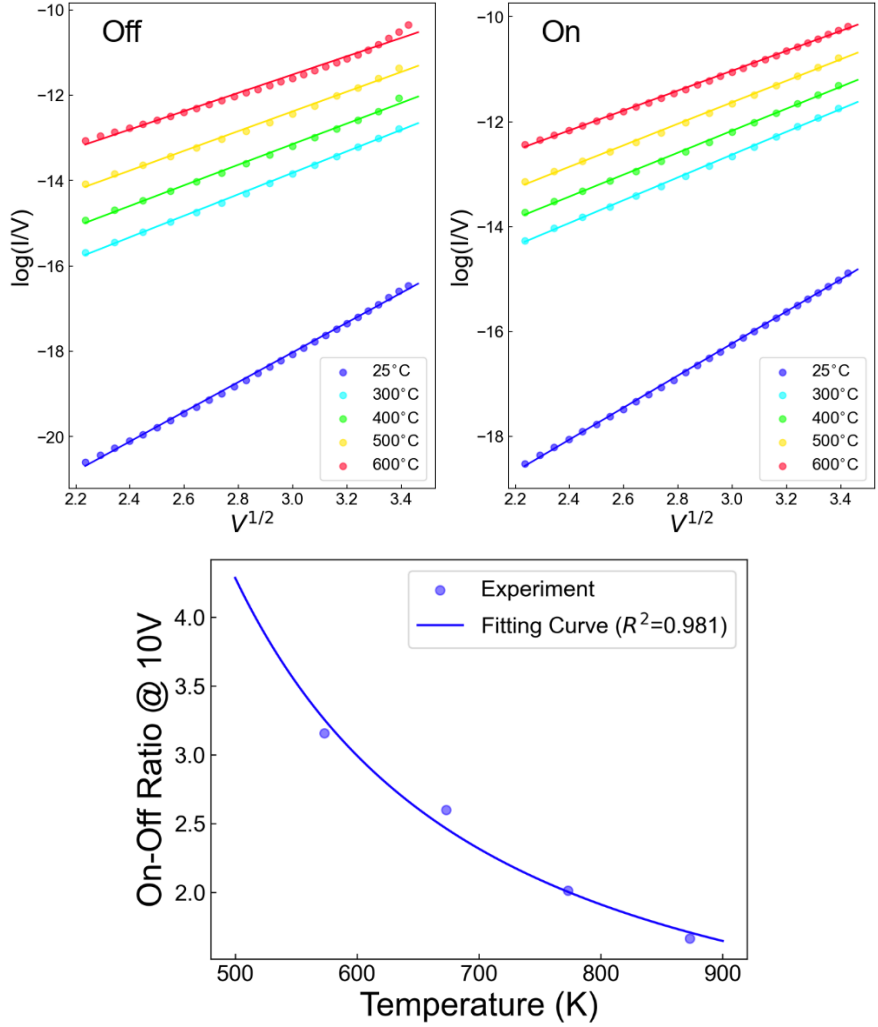

Fig. S6. (Upper panel) Poole-Frenkel model fitting to HRS and LRS states of the device at various temperatures. (Lower Panel) Fitting of our On-Off ratio temperature dependence model to the experimental data.

As demonstrated in Figure S6 (upper panel), both HRS and LRS I-V characteristics of our AlScN ferroelectric diode follow the Poole-Frenkel (P-F) emission model with the following expression for current density:

$$J \propto E \, exp\left(\frac{-q\phi_B + \sqrt{qE/(\pi\epsilon)}}{kT}\right)$$

An On-Off ratio is created as the internal electric field $E$ is modulated based on the polarization of the ferroelectric material. Taking $E_1$ and $E_2$ as the internal electric field at a fixed applied voltage for the HRS and LRS states of the device respectively, we can then write the On-Off ratio as the following:

$$\text{On-Off Ratio} = R = \frac{E_2}{E_1} \exp\left[\frac{1}{kT}\left(\sqrt{\frac{qE_2}{\pi\epsilon}} - \sqrt{\frac{qE_1}{\pi\epsilon}}\right)\right]$$

This expression allows us to clearly deduce that the On-Off ratio decreases with temperature according to the relation $R \propto exp\left(\frac{\alpha}{kT}\right)$ where $\alpha = \left(\sqrt{\frac{qE_2}{\pi\epsilon}} - \sqrt{\frac{qE_1}{\pi\epsilon}}\right)$. Our experimental data demonstrates an excellent fit ($R^2=0.981$) to this model in the 300-600 degrees Celsius regime as shown in lower panel of figure S6. We note that the possible presence of other limiting conduction mechanisms (e.g. thermionic emission) in the Off state I-V characteristics at room temperature. As such, it is not included in the fit to our model.

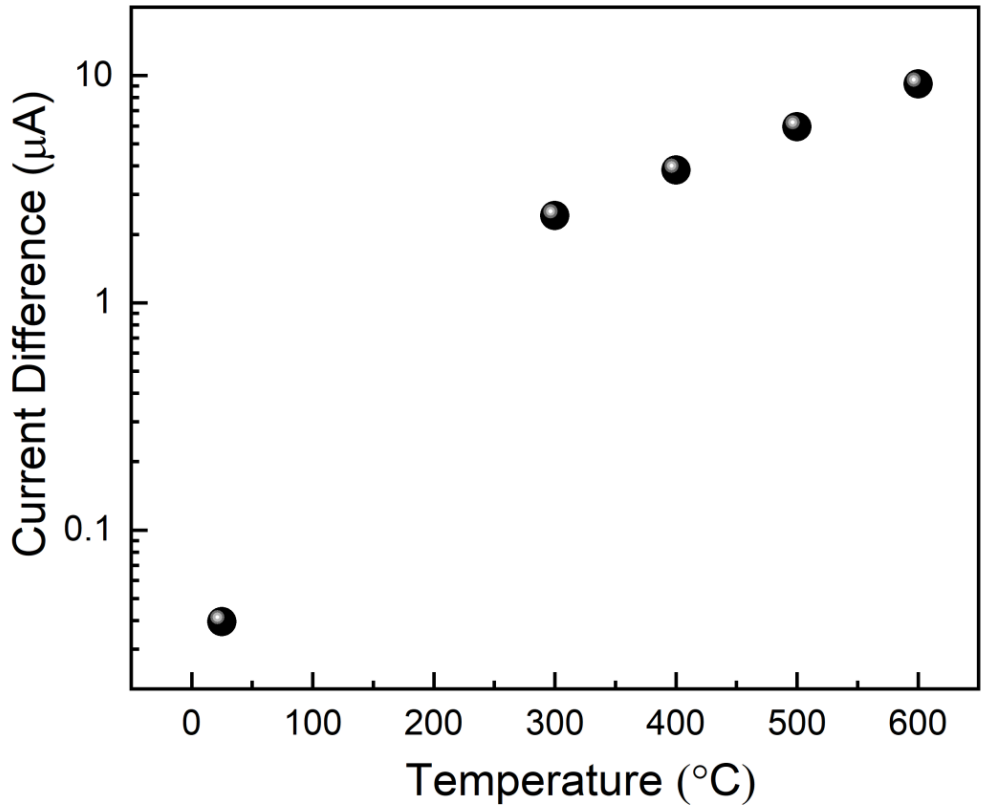

Fig. S7. Temperature dependence of absolute current difference.

We observed the decrease of On-Off ratio with increasing temperature, whereas the current difference (On current- Off current) was found to increase with increasing temperature. This indicates that the reduction in On-Off ratio is due to a larger global increase in leakage current.

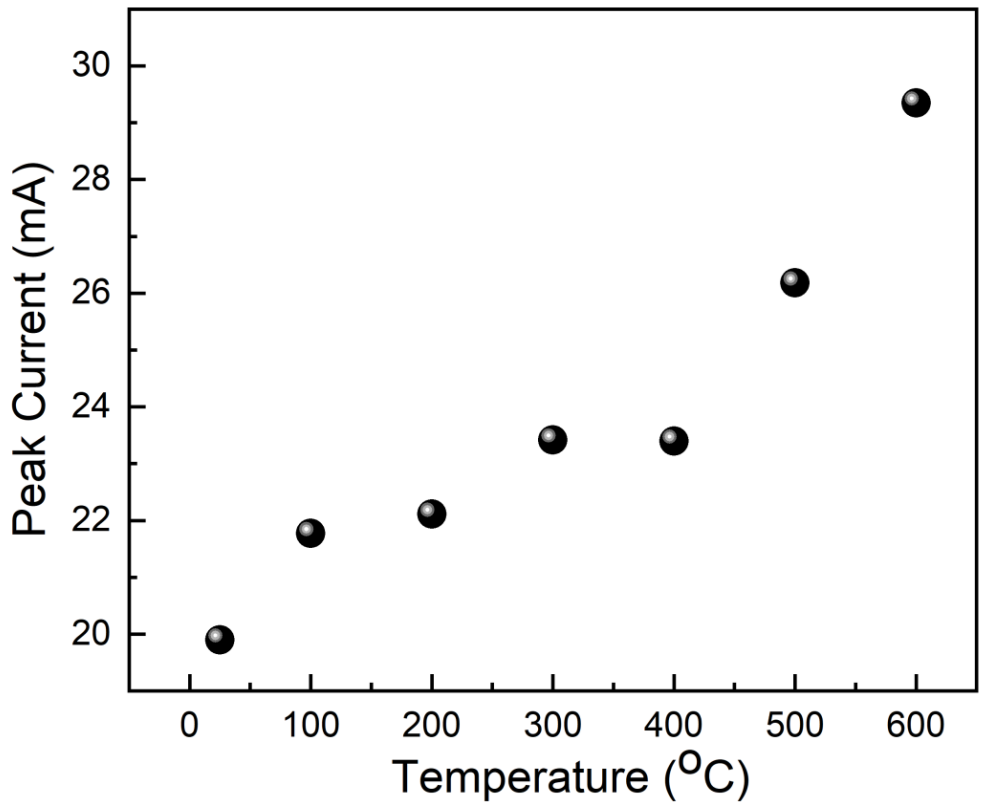

Fig. S8. Temperature dependence of Peak Current.

The peak current obtained from J-t curve (Fig.3 (b)) increases with increase of temperature due to the decrease in available switching time as the switching voltage deceases as well as increase in overall leakage current arising from increasing thermal energy which reduces activation barrier for hopping or defect assisted tunneling transport.

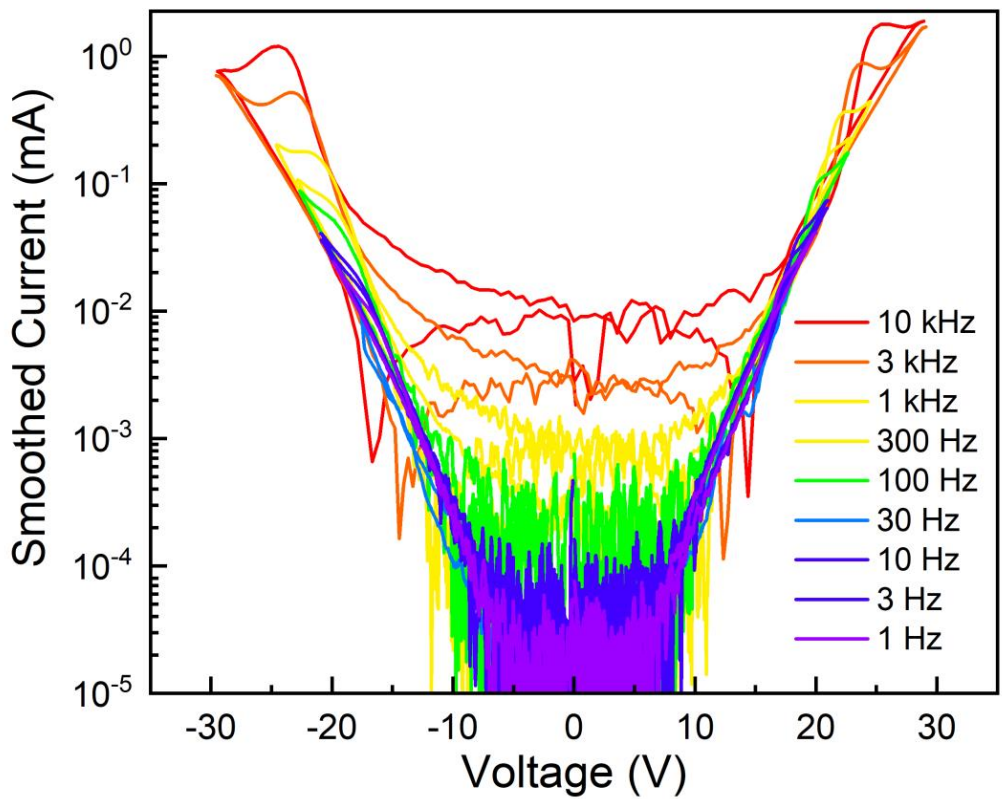

Fig. S9: Pulsed I-V rates spanning from 1Hz to 10 kHz.
As ramp rate decreases, the switching current spreads out over the course of the pulse and becomes dominated by the leakage current. By 1Hz, the Pulsed-IV is almost identical to the DC-IV. Note that leakage current is a DC effect and is therefore not affected significantly by ramp rate, however higher ramp rates can survive higher voltages.

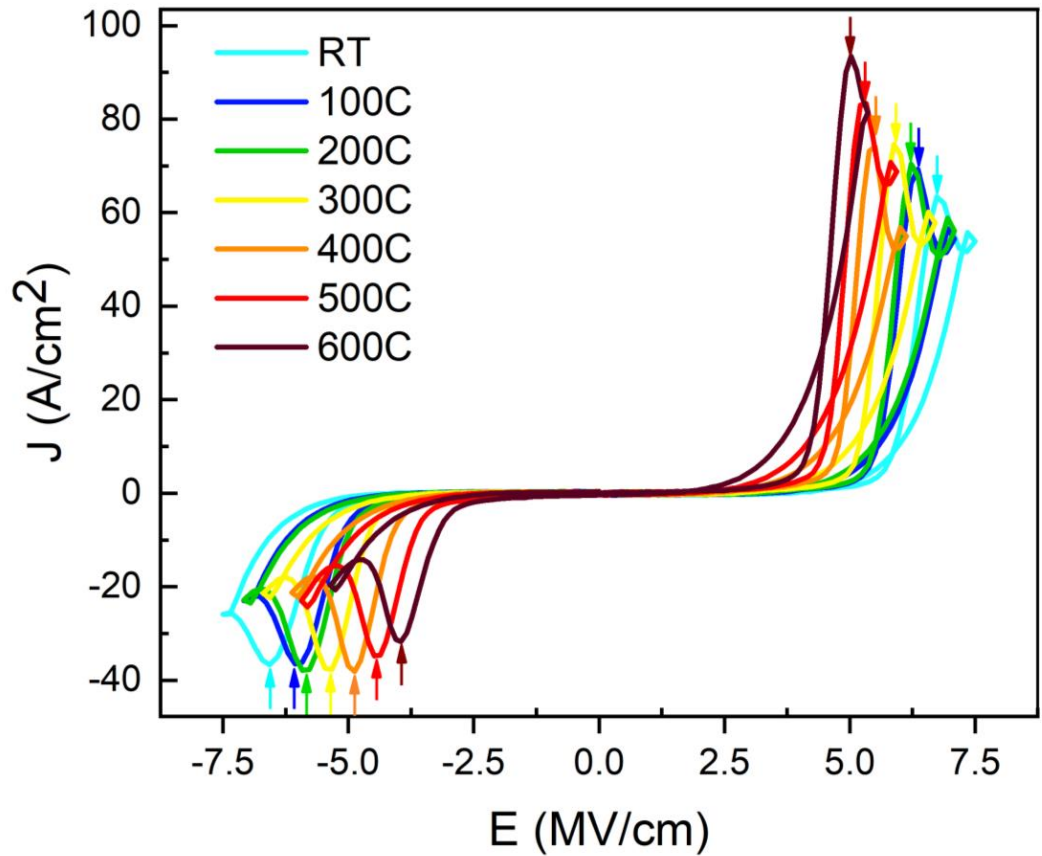

Fig. S10: Pulsed I-V at different temperatures indicating the coercive field.
At high ramp rates, the switching current creates a dominant peak in the IV curve. This peak clearly defines a switching voltage from which the coercive field was calculated. This value moves towards zero with increasing temperature as with DC-IV.

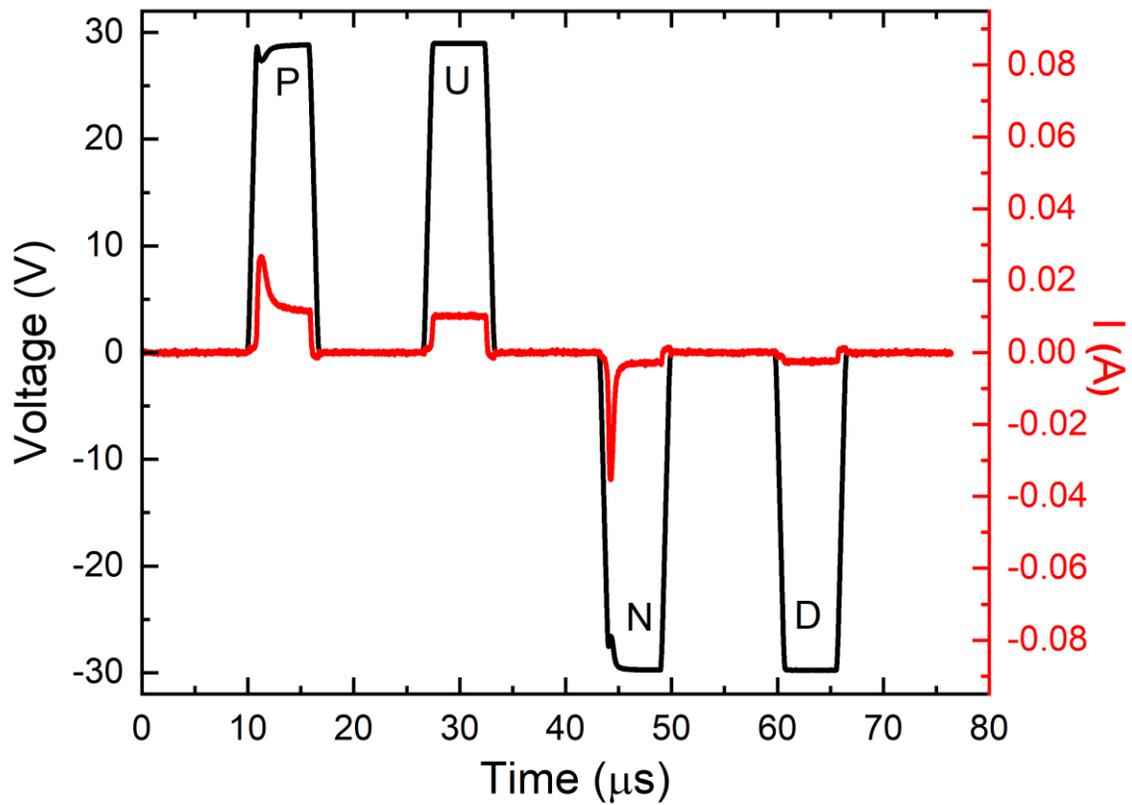

Fig. S11. PUND measurement performed at RT on a 100 µm diameter of top Ni electrode. The Positive (P), up (U), negative (N) voltage pulses are shown in the figure.
The MIM devices clearly demonstrate ferroelectric switching for both positive and negative voltage pulses at RT.

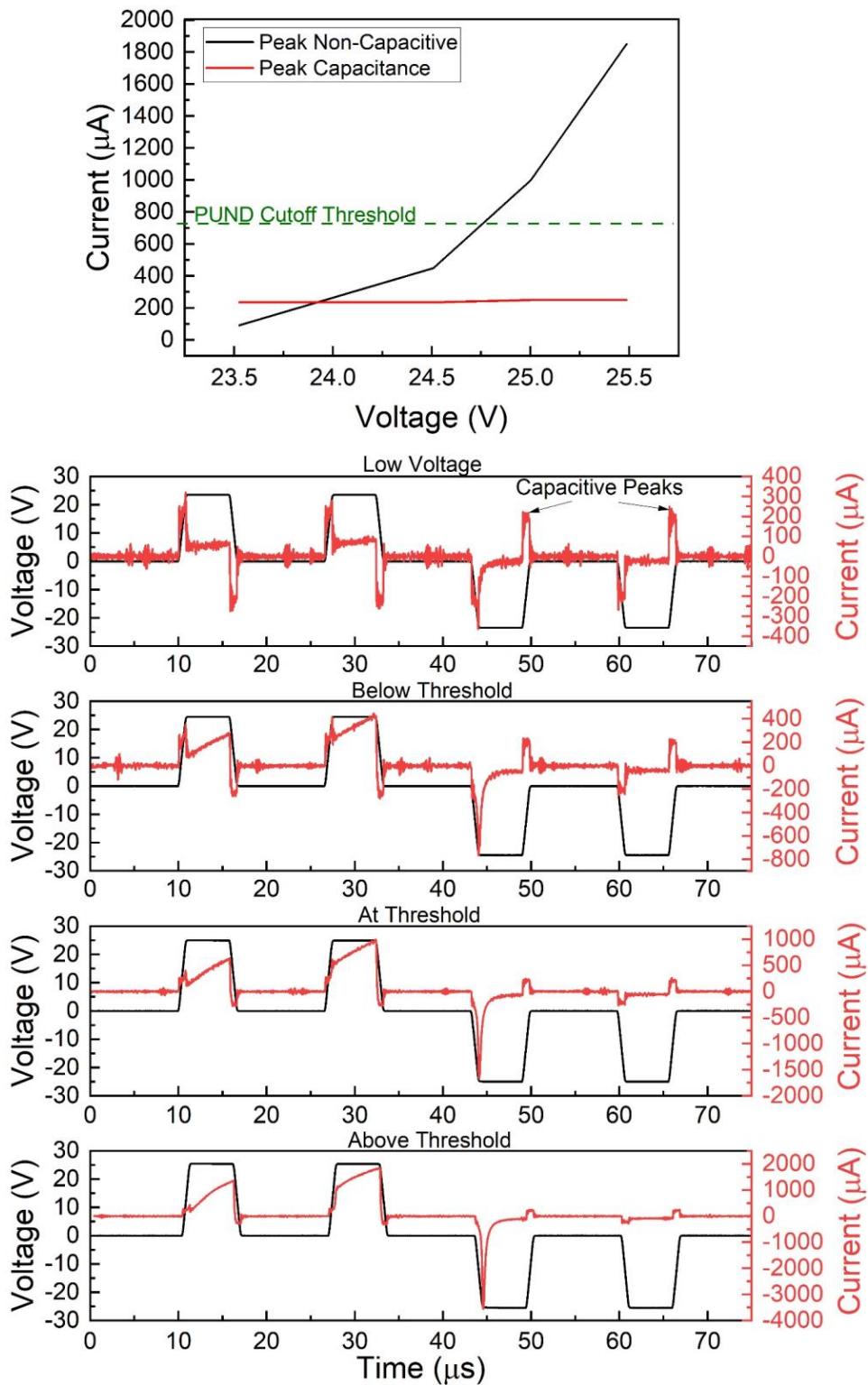

Fig. S12. PUND with capacitive current and onset level. (Left) Select PUND measurements at room temperature showing capacitive current and increasing leakage current. (Right) Non-capacitive (leakage+switching) and capacitive currents plotted vs voltage.

Because the "onset" value of the diode response was not independently measured, it was selected from lower PUND measurements. In general, the PUND maximum started at 5 V and was walked up by 500 mV increments. At low voltages, no pulse or leakage current is evident, and instead only a charging spike owing to capacitance (both device and parasitic) is observed. At diode onset, this charging is rapidly dominated, and this point was used to define diode onset. The non-capacitive current rises quite rapidly at the onset of switching, and thus the onset threshold is not very sensitive to the selected ratio.

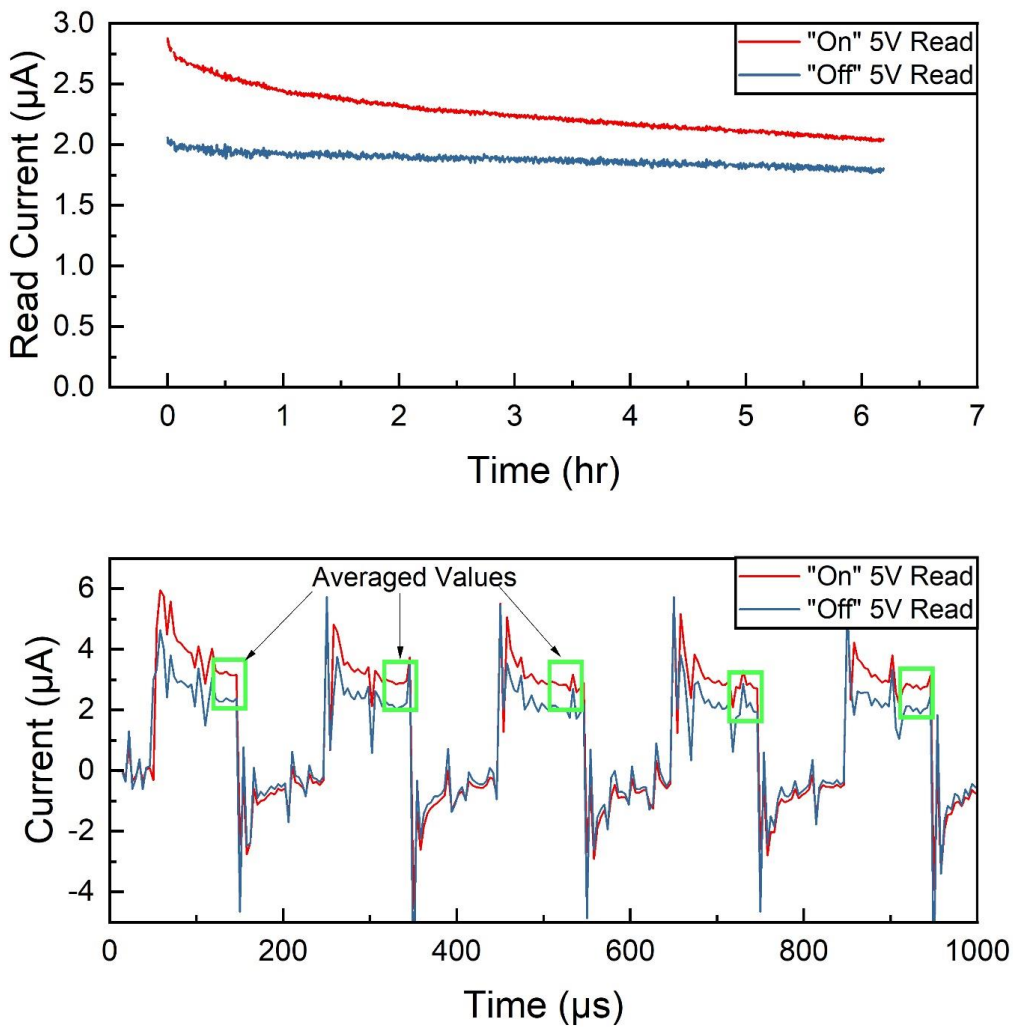

Fig. S13. (Top) Reproduction of Fig. 4b. (Bottom) Raw pulse currents for read tests demonstrating averaging scheme.

The 6-hour On-Off ratio was measured via groups of 25 pulses every 15 seconds. The pulses were widened to 100 µs to avoid an instrumentational parasitic capacitance and resistance that made an obvious charging current for the first 30-50 µs or so. The last few points of all these pulses were averaged to produce a single data point. Note that 25 pulses were averaged for each data point, but only 5 are shown for clarity. Current noise is similar on both read traces because the reads were simultaneous, indicating instrumentational rather than device noise as the cause.

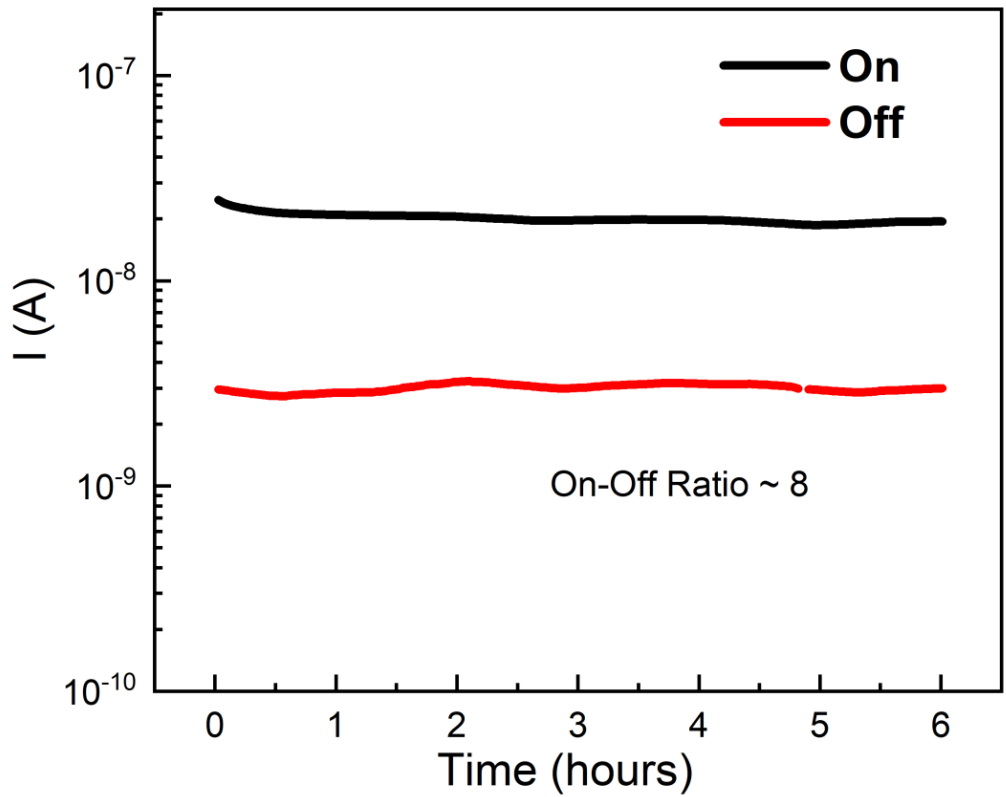

Fig. S14. Retention behavior at 5 V read voltage at RT.
At room temperature, our AlScN MIM device demonstrated a stable retention for over 6 hour with a stable On-Off ratio of ~8.

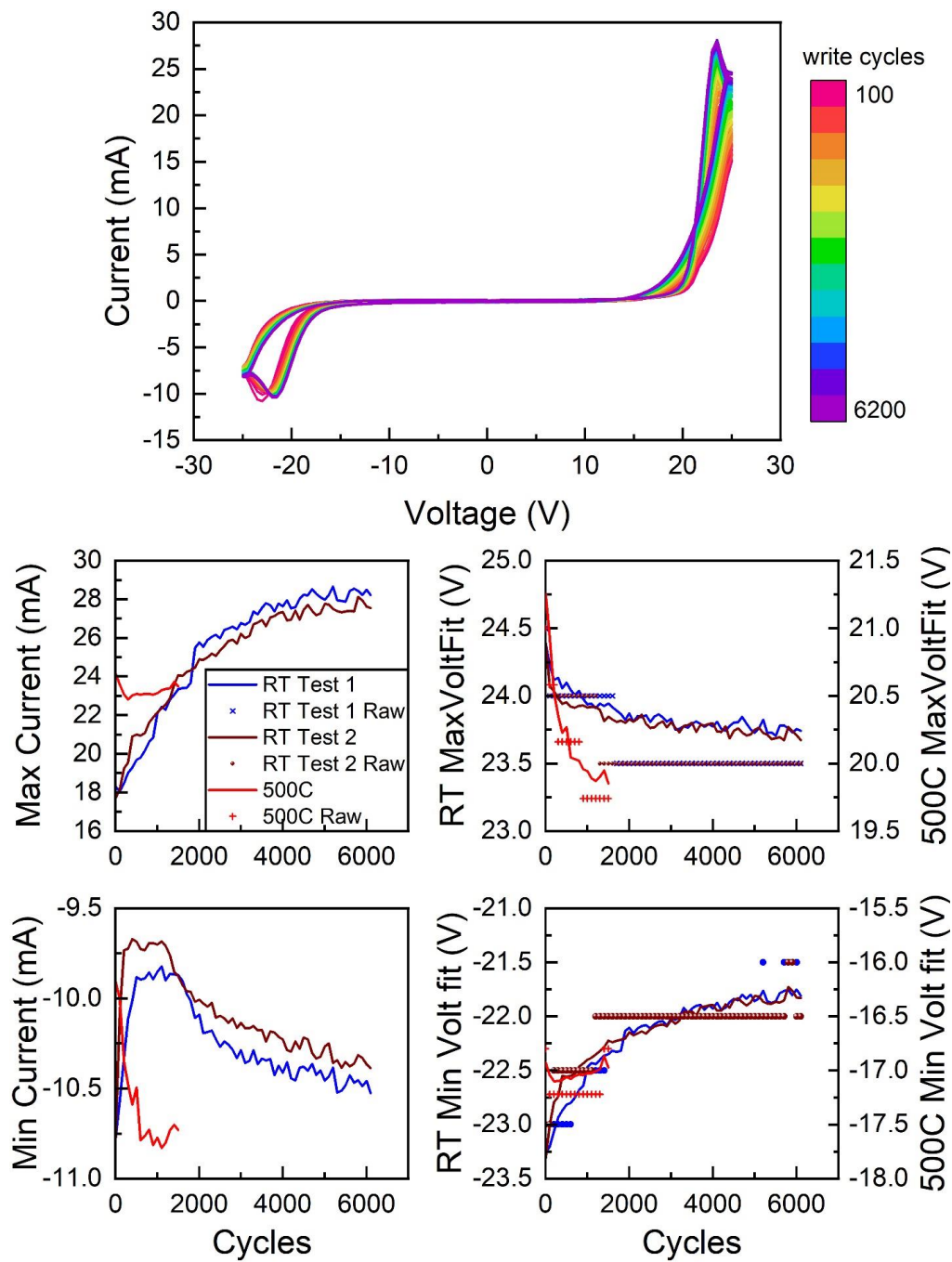

Fig. S15. (Top) Write endurance test at performed at RT. (bottom) comparison of peak trends in RT tests vs. 500 °C.

Write cycles performed at room temperature similar to Fig. 4(c). What is displayed are test Pulsed-IV cycles in between sets of 100 write pulses to determine the health of the device. Evolution of the peak heights and locations is shown below. "VoltFit" refers to a parabolic fit of the top three data points near the IV peak, since the actual data resolution (shown as "Raw") was too low to see meaningful trends.

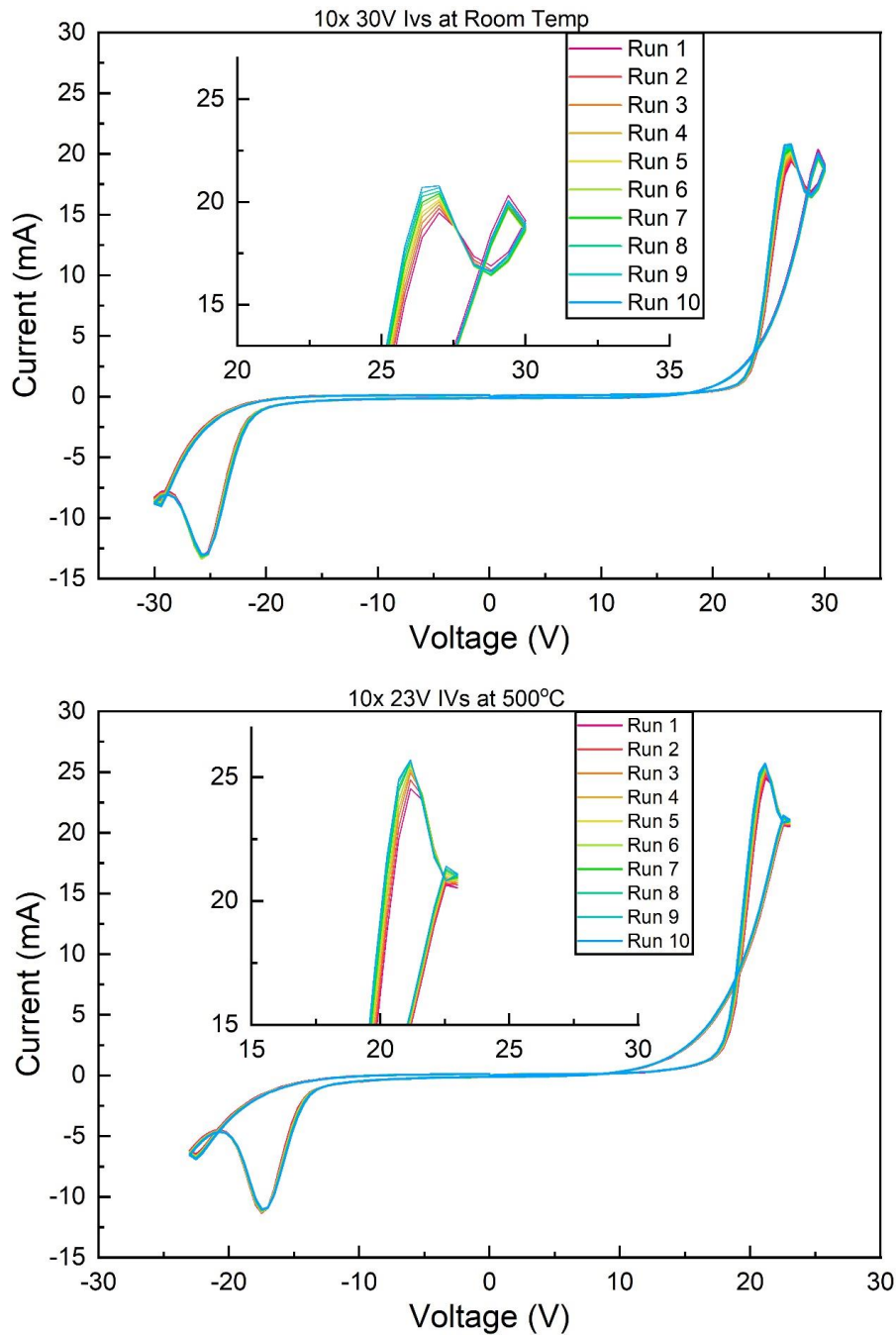

Fig. S16. Wake-up effect observed over first ten cycles observed at (top) room temperature (bottom) 500 °C.

Device wake-up is a well-known phenomenon whereby depinning and domain wall motion of the ferroelectric domains occurs most notably in the first few switches. This is similar in mechanism to the observed changes seen in Fig. S15, but here the device has not been stressed prior to measurement. The wake-up effect in the first ten cycles at both room temperature and 500 °C are modest.